\newif\ifAnon\Anonfalse
\newif\ifDraft\Draftfalse
\newif\ifFinal\Finalfalse

\ifFinal\Anonfalse\Draftfalse\fi
\documentclass[runningheads]{llncs}
\usepackage[T1]{fontenc}

\usepackage[htt]{hyphenat} %
\usepackage{hyperref}
\usepackage{cispa_defaults}
\usepackage{graphicx}
\usepackage{makecell}
\usepackage{multirow}
\usepackage{breakcites} %
\usepackage[shortcuts]{extdash} %

\newcommand{\parhead}[1]{\textbf{#1}.\ }
\newcommand{\code}[1]{\texttt{#1}\xspace}
\newcommand{\meltdownthreea}{Meltdown\-/CPL\-/REG\xspace}
\newcommand{\meltdownus}{Meltdown\-/US\-/L1\xspace}

\newcommand{\Microcode}{$\mu$code\xspace}

\newcommand{\PerfAttack}{CounterLeak\xspace}
\newcommand{\AnalysisTool}{RegCheck\xspace}

\newcommand{\LabOneCPU}{Intel Celeron J4005\xspace}
\newcommand{\LabOneOS}{Ubuntu 20.04\xspace}
\newcommand{\LabOneKernel}{Linux kernel 5.4.0\xspace}

\newcommand{\LabFifteenCPU}{Intel Celeron N3350\xspace}
\newcommand{\LabFifteenOS}{Ubuntu 22.04\xspace}
\newcommand{\LabFifteenKernel}{Linux kernel 5.15.0\xspace}

\begin{document}
\title{Reviving Meltdown 3a}

\author{Daniel Weber \and
Fabian Thomas \and
Lukas Gerlach \and\\
Ruiyi Zhang \and
Michael Schwarz}

\authorrunning{Weber et al.}

\institute{CISPA Helmholtz Center for Information Security\\
Saarbr\"ucken, Saarland, Germany\\
\email{<firstname>.<lastname>@cispa.de}}

\maketitle

\begin{abstract}
Since the initial discovery of Meltdown and Spectre in 2017, different variants of these attacks have been discovered.
One often overlooked variant is Meltdown 3a, also known as \meltdownthreea.
Even though \meltdownthreea was initially discovered in 2018, the available information regarding the vulnerability is still sparse.

In this paper, we analyze \meltdownthreea on 19 different CPUs from different vendors using an automated tool.
We observe that the impact is more diverse than documented and differs from CPU to CPU.
Surprisingly, while the newest Intel CPUs do not seem affected by \meltdownthreea, the newest available AMD CPUs (Zen3+) are still affected by the vulnerability.
Furthermore, given our attack primitive \PerfAttack, we show that besides up-to-date patches, \meltdownthreea can still be exploited as we reenable performance-counter-based attacks on cryptographic algorithms, break KASLR, and mount Spectre attacks. 
Although \meltdownthreea is not as powerful as other transient-execution attacks, its attack surface should not be underestimated.
\end{abstract}

\section{Introduction}\label{sec:intro}

Microarchitectural side-channel attacks have been known for several decades~\cite{Kocher1996}.
These attacks exploit the side effects of CPU implementations to infer metadata about actual data being processed by the CPU.
Well-known examples of microarchitectural side-channel attacks include cache attacks, \eg \FlushReload~\cite{Yarom2014Flush} or \PrimeProbe~\cite{Percival2005}, which have been used to leak cryptographic secrets~\cite{Aciicmez2008,Yarom2014Flush,Liu2015Last} or violate the privacy of users, \eg by spying on user input~\cite{Oren2015,Gruss2015Template,Lipp2016,Schwarz2018KeyDrown}.
Another example of side-channel attacks are attacks based on the CPUs performance counters~\cite{Uhsadel2008,Dixon2017,Bhattacharya2015}.
However, these attacks are considered mitigated as access to performance counters is restricted on modern CPUs~\cite{Dixon2017}.

In 2017, transient execution attacks were first discovered in the form of Meltdown~\cite{Lipp2018meltdown} and Spectre~\cite{Kocher2019}.
Shortly afterward, a variety of transient execution attacks were discovered~\cite{Canella2019A,Schwarz2019ZL,VanSchaik2019RIDL,Weisse2018foreshadow,Maisuradze2018spectre5,Koruyeh2018spectre5,Ragab2021crosstalk,Ragab2021rage}.
One attack that is often considered less powerful than other variations, and thus easily overshadowed by the discovery of other variants, is Meltdown 3a~\cite{Canella2019A,ARMSpecAnalysis_whitepaper}, later on, referred to as \meltdownthreea in the extended transient-execution attack classification by Canella~\etal\cite{Canella2019A}.
\meltdownthreea allows an unprivileged attacker to leak the content of system registers restricted to privileged access.
After the discovery of the attack, CPU vendors reacted with microcode updates to fix the vulnerabilities~\cite{Intel2023Affected, ARMSpecAnalysis_whitepaper}.
More precisely, CPU vendors fixed the vulnerability for system registers containing confidential information, such as model-specific registers.

In this paper, we show that \meltdownthreea exposes a more complex attack surface than originally thought, which allows an attacker to exploit it, even 5 years after the initial discovery of the attack.
Although the Meltdown variant itself is known, there is no systematic analysis yet.
Thus, we introduce \AnalysisTool, an automated tool to test x86 CPUs for various \meltdownthreea variants.
Our analysis using \AnalysisTool reveals two main insights.
First, CPUs that are vulnerable to \meltdownthreea do not show the same leakage for all system registers.
Instead, the analysis shows that different CPUs expose leakage of different system registers.
Hence, the category \meltdownthreea is too coarse-grained to determine if a CPU is affected.
The official tables published by Intel~\cite{Intel2023Affected} comment only on the leakage of the \code{rdmsr} instruction.
Nevertheless, \AnalysisTool shows that for some of these CPUs, there is at least one system register that can be leaked. 
Second, the fact that a CPU is unaffected by the original Meltdown attack, \ie \meltdownus~\cite{Lipp2018meltdown,Canella2019A}, does not imply that the CPU is also unaffected by \meltdownthreea as we observe leakage until the newest tested AMD CPUs.
Our analysis shows that while \meltdownthreea was mitigated using microcode updates for system registers containing confidential data, \meltdownthreea is still possible on modern CPUs for those privileged registers that are not considered confidential, including registers containing only metadata about a program, such as performance counters.

Based on these observations, we introduce the attack primitive \PerfAttack.
\PerfAttack allows unprivileged attackers to read performance counters, thereby leaking performance monitoring metadata about applications running on a system.
This shows that the state-of-the-art \meltdownthreea mitigations are insufficient for protecting against side-channel leakage.
In our proof-of-concept attack, we read the performance counters to leak meta information about applications.
We encode transiently-read data in the form of a Spectre attack with \SI{66.7}{\bit/\second}, but with a generic encoding gadget. %
We also break the security mitigation Kernel Address Space Layout Randomization (KASLR) by leaking meta information of the page-table walker when accessing potential kernel pages.
Furthermore, \PerfAttack re-enables attacks that rely on performance counters~\cite{Bhattacharya2017perf,Aciicmez2007predicting}.
These attacks are considered mitigated because the required performance interface was made privileged.
Our attack extracts an RSA key from a square-and-multiply implementation based on MbedTLS.
We demonstrate a full key recovery of a 2048-bit key within \SI{15}{\minute}.
We also show that \PerfAttack can be used to break the Zigzagger branch-shadowing mitigation~\cite{Lee2017Inferring}.
While all these attacks require that the underlying system has performance counters enabled, this is the case for various performance-counter-based defenses that were proposed~\cite{Zhang2013,Irazoqui2018mascat,Payer2016,Zhang2016CloudRadar,Chiappetta2015,Mushtaq2020,Mushtaq2018nightswatch,Wang2020hybrid,Wang2020comprehensive,Zhang2021see}.
Thus, we stress that when designing defense tools, it is crucial to evaluate the additional attack surface introduced by these tools.

To summarize, we make the following contributions:
\begin{compactenum}
    \item We analyze 19 CPUs of different vendors using an automated tool, showing that \meltdownthreea was never fully mitigated and can still be exploited.
    The analysis tool is open-source and can be found on GitHub\footnote{\url{https://github.com/cispa/regcheck}}.
    \item We use our side channel for a novel Spectre attack using performance counters and to bypass KASLR based on the performance characteristics of the page-table walker.
    \item We re-enable attacks on cryptographic libraries.
\end{compactenum}

\textbf{Outline.}
\Cref{sec:background} provides background.
\Cref{sec:meltdown-analysis} discusses our analysis of \meltdownthreea across different Intel and AMD CPUs.
\Cref{sec:primitive} presents the \PerfAttack primitive, and \Cref{sec:eval} evaluates the primitive.
\Cref{sec:case} shows 4 case studies based on the attack primitive.
\Cref{sec:countermeasures} discusses mitigations to prevent the exploitation of \meltdownthreea and \PerfAttack.
\Cref{sec:discussion} discusses related work and the generalization of our insights.
\Cref{sec:conclusion} concludes.

\textbf{Responsible Disclosure.}
We disclosed our findings to Intel on February 15, 2023 and AMD on February 16, 2023.
While both vendors got back to us, neither plan to roll out mitigations for the new findings.

\section{Background} \label{sec:background}
In this section, we provide the background for this paper. 
We introduce performance counters as we attack this interface in the remainder of the paper.
We introduce side channels and transient-execution attacks, as these concepts are crucial for the understanding of our attack implementation.

\subsection{Performance Counters}\label{subsec:background-perf}

Modern CPUs expose performance counters to help developers analyze and benchmark their programs.
Performance counters keep track of different microarchitectural events, such as the number of issued micro-operations or the number of evicted cache lines from the L1D cache.
Performance counters are programmed to record a specific event.
The current count of the event can be read using the x86 instruction \instr{rdpmc}.
The privilege level needed to execute \instr{rdpmc} can be configured by the operating system.
For example, Linux exposes this configuration via the file \code{/sys/devices/cpu/rdpmc}.
In the past, unprivileged access to performance counters was exploited to mount side-channel attacks and break KASLR~\cite{Dixon2017,Uhsadel2008}.
Thus, modern operating systems, such as Debian 11, Ubuntu 20.04, or Fedora 35, disallow the access to the performance monitoring interface.

\subsection{Side Channels}
The term side channel refers to a meta information leaking from a system that can be used to reason about the actual inaccessible data being processed by the system.
In CPU microarchitectures, this meta information occurs in various forms, including power usage~\cite{Lipp2020Platypus}, access timings~\cite{Percival2005,Gruss2016Flush}, and contention~\cite{Aldaya2018,Gras2020,paccagnella2021lord}.
An attack exploiting observable meta information is referred to as a side-channel attack.
Microarchitectural software-based side-channel attacks (in the remainder of this paper just referred to as ``side-channel attacks'') have been demonstrated against cryptographic algorithms and libraries~\cite{Percival2005,Lou2021survey,Yarom2014Flush,Bhattacharya2015}, to spy on users~\cite{Schwarz2018KeyDrown}, and to break security boundaries~\cite{Frisk2016KASLR,Gruss2017KASLR,Dixon2017}.
Over the last few decades, researchers have demonstrated side-channel attacks based on several microarchitectural components, such as the CPU caches~\cite{Percival2005,Yarom2014Flush,Gruss2016Flush,Purnal2021PrimeScope}, the execution units~\cite{Aldaya2018,Rokicki2022port}, or the component's power consumption~\cite{Lipp2020Platypus}.

\subsection{Transient-Execution Attacks}
Transient-execution attacks exploit performance optimizations of the microarchitecture. They are split into two major categories, namely Meltdown-type and Spectre-type attacks, based on the type of performance optimization they exploit~\cite{Canella2019A,Intel2020Refined}.
While Spectre-type attacks exploit branch predictors, Meltdown-type attacks exploit faulting instructions for which the processor continues to execute depending instructions.
These instructions can compute with the values of the faulting instructions until the fault is recognized by the CPU and the instruction stream is rolled back to before the faulting instruction.
These instructions that were executed but never architecturally visible because of the roll-back, are called transient instructions~\cite{Canella2019A,Intel2020Refined}.
One Meltdown-type attack that is typically considered less critical, is \meltdownthreea (initially called Meltdown 3a)~\cite{ARMSpecAnalysis_whitepaper,IntelrogueSystemRegRead2018AffectedRegisters,Canella2019A}.
\meltdownthreea allows an unprivileged attacker to leak the content of privileged system registers.
Hereby, the attacker reads the system registers via a designated instruction such as \instr{rdmsr} and encodes the content into a microarchitectural element before the roll-back occurs.
Afterward, the attacker can decode this information using a side-channel attack, thus leaking the system register's content.
To mitigate the impact of \meltdownthreea, CPU vendors provide microcode updates for affected systems~\cite{Intel2023Affected,ARMSpecAnalysis_whitepaper}.

\section{Analysis of \meltdownthreea}\label{sec:meltdown-analysis}

For \meltdownthreea, microcode prevents the leakage of system registers containing sensitive values. 
However, other registers containing meta-data about applications can still be leaked, enabling another source of side-channel leakage.
We present the first systematic analysis of \meltdownthreea~\cite{ARMSpecAnalysis_whitepaper,IntelrogueSystemRegRead2018AffectedRegisters,Canella2019A} to analyze the remaining attack surface after applying state-of-the-art microcode patches. %
To systematically analyze CPUs, we design \AnalysisTool to test a CPU for different \meltdownthreea variants automatically.
Our analysis of 19 systems leads to two main insights.
First, if a system is vulnerable to \meltdownthreea, this does not mean that \textit{all} system registers are affected.
Second, even fully patched recent CPUs unaffected by the original Meltdown attack (\meltdownus)~\cite{Lipp2018meltdown} can be vulnerable to \meltdownthreea.

\parhead{Design and Implementation}
Our prototype of \AnalysisTool is developed for Intel and AMD CPUs running Linux.
Note that the same approach can be ported to other architectures, \eg to support Arm CPUs, as this is purely an engineering task.
\AnalysisTool tests a list of different system registers that are either only accessible for privileged users or can be configured to only allow privileged access.
The list is based on Intel's list of affected registers~\cite{IntelrogueSystemRegRead2018AffectedRegisters}.
We provide a complete list of analyzed system registers in \Cref{tab:sysreg-tab}.
The inner workings of \AnalysisTool can be broken down into two steps:

First, \AnalysisTool changes the kernel parameters to a consistent state for the measurements.
More precisely, one CPU core is isolated using the \code{isolcpus} kernel parameter, and unprivileged access to \code{rdfsbase} and \code{rdgsbase} is disabled using the \code{nofsgsbase} kernel parameter.
Similarly, the access to further system registers which are not permanently restricted to privileged access, \eg performance counters (\cf \Cref{subsec:background-perf}), is configured to prevent unprivileged access to these registers before testing.
After applying these settings, \AnalysisTool executes on the isolated CPU core to reduce the system noise for its measurements.
Next, for each system register, \AnalysisTool tries to reason about its exploitability.
To do so, \AnalysisTool tries to exploit \meltdownthreea and encode 8 bits of the system register into a lookup array.
The encoding is done by transiently accessing the corresponding index of the array, \eg if the leaked bits form the value 7, then an access to \code{array[7 * N]} is performed. %
The resulting fault can either be suppressed or handled.
For \AnalysisTool we choose to handle the fault using a signal handler as this approach is portable to all modern CPUs.
Our implementation varies $N$ from 1024 to 4096 bytes to find a good tradeoff between the size of memory pages needed to encode the values while still preventing different accessing from either directly going into the same cache line or prefetching other array entries.
Note that we choose to encode 8 bits instead of only 1 bit to distinguish actual leakage from system noise better.
After encoding these bits, the tool checks whether a transient access to any index has taken place by iterating over the array and performing \FlushReload, \ie timing the memory access to each array index.
If \AnalysisTool succeeds at leaking the target system register multiple times, it flags it as vulnerable. %
We test our tool on Intel and AMD CPUs from different generations.
All tests use the latest microcode available in the Ubuntu repositories.
For further details on the specific microcode version used we refer the reader to \Cref{tab:meltdownthreea-analysis-overview}.%

\parhead{Affected Registers}
The main insight from our analysis is that not all privileged registers are affected in the same way by \meltdownthreea.
This is especially interesting because Intel's list of CPUs affected by certain vulnerabilities~\cite{Intel2023Affected} (accessed May 2023) only lists CPUs where the \code{rdmsr} instruction can be exploited by \meltdownthreea.
However, our results in \Cref{tab:meltdownthreea-analysis-overview} show that some CPUs that Intel flags as unaffected by the \meltdownthreea \code{rdmsr} leakage can still be exploited to leak the contents of other system registers, such as the performance counters using \code{rdpmc}.
This, for example, is the case for the Intel Celeron J4005 and the Intel Celeron N3350.
The results in \Cref{tab:meltdownthreea-analysis-overview} show that the instruction \instr{rdfsbase} leaks on 8 out of 14 CPUs affected by \meltdownthreea.
The CPU timestamp counter accessed via \instr{rdtsc} or \instr{rdtscp} leaks on 2 out of 14 affected CPUs.
Performance counter leak on 3 of the affected CPUs via \instr{rdpmc}.
A possible explanation for these different leakage rates could be that for executing \instr{rdpmc}, the CPU has to decode an argument of the instruction, \ie the index of the access performance counter stored in \instr{RCX}, while for \instr{rdtsc}, \instr{rdtscp}, and \instr{rdfsbase} all required information to fetch the requested data is available, leading to a potentially simpler execution path.
Nevertheless, the CPUs where \instr{rdpmc} is vulnerable do not show a superset of the vulnerable instructions compared to the other systems.
Even though these systems show vulnerable \instr{rdpmc} implementations, we could not verify further leakage.

\parhead{Affected CPUs}
Our second insight is that the fact that CPUs are vulnerable to \meltdownus is not related to whether a CPU is also vulnerable to \meltdownthreea, as shown in \Cref{tab:meltdownthreea-analysis-overview}.
In other words, we can leak from system registers of CPUs that are affected by \meltdownus and of CPUs not affected by \meltdownus.
This is especially surprising for recently released CPUs, such as the Ryzen 9 6900HX.
We observe that the tested Intel CPUs from Alder Lake onward do not show leakage, while newer AMD CPUs do.

\parhead{\AnalysisTool Limitations}
The current proof-of-concept implementation of our tool \AnalysisTool comes with different limitations.
We do not check for the leakage of \instr{swapgs} as previous work has already analyzed this instruction and its leakage potential~\cite{Lutas2019}.
We neither check the \instr{xgetbv} instruction.
The reason for the latter is that to prevent unprivileged access to \instr{xgetbv}, \AnalysisTool needs to set the \code{OSXSAVE} bit of \instr{CR4}, which crashes the tested OS.
A detailed list of the analyzed system registers is shown in \Cref{tab:sysreg-tab}.

\begin{table}[tb]
    \caption{System registers and their access instructions tested by \AnalysisTool.}
    \label{tab:sysreg-tab}
    \centering
    \adjustbox{max width=\hsize}{
        \begin{tabular}{ll}
            \toprule
            \textbf{Access Instruction~} & \textbf{Details}\\
            \midrule
            \instr{rdpmc} & Reads the specified Performance counter \\
            \instr{rdtsc} & Reads the CPU timestamp counter \\
            \instr{rdtscp} & Reads the CPU timestamp counter \\
            \instr{mov CRx} & Loads the Control registers 0 - 8\\
            \instr{mov DRx} & Loads the Debug registers 0 - 7 \\
            \instr{rdfsbase} & Retrieves segment selector of the FS segment base register \\
            \instr{rdgsbase} & Retrieves segment selector of the GS segment base register \\
            \instr{rdmsr} & Model Specific Registers \\
            \instr{str} & Loads the segment selector of the Task register \\
            \instr{sldt} & Loads the segment selector from the Local Descriptor Table register \\
            \instr{sidt} & Loads the segment selector from the Interrupt Descriptor Table register \\
            \instr{sgdt} & Loads the segment selector from the Global Descriptor Table register \\
            \instr{smsw} & Loads the Machine status word \\
            \bottomrule
        \end{tabular}
    }
\end{table}

\Cref{tab:meltdownthreea-analysis-overview} flags \instr{rdtsc} and \instr{rdtscp} for certain instances with an ``U'' (short for ``unverified'').
On these systems, we observed leakage from the system registers, but could not verify that the leakage stems from the CPU timestamp counter.
The reason for this is that \AnalysisTool uses a counting thread as a timer for analyzing the instructions \instr{rdtsc} and \instr{rdtscp}.
However, this timer does not work reliably on CPUs not supporting hyperthreading, as the counting and attacker thread yield a more accurate timer when both threads execute on co-located hyperthreads.
\Cref{tab:meltdownthreea-analysis-overview} also has system registers flagged with ``ZF'' (short for ``zero forwarding'').
For these registers, an access always returns the value 0 instead of the actual value.
While such behavior intuitively sounds invulnerable, instructions forwarding zero values already led to microarchitectural attacks~\cite{Vanbulck2020lvi,Canella2020kaslr}.

\begin{table}[bt]
    \caption{CPUs tested by \AnalysisTool for \meltdownthreea. ``U'' means we could not verify if an actual timestamp is leaked. ``ZF'' means that only the value 0 is returned transiently. Additionally, we annotate machines that are vulnerable to the original Meltdown attack.}
    \label{tab:meltdownthreea-analysis-overview}
    \centering
    \adjustbox{max width=\hsize}{
        \begin{tabular}{lllrcl}
            \toprule
            \textbf{CPU}   & \textbf{\Microcode} & \textbf{$\mu$arch} & \textbf{Release~} & \textbf{MD-US~} & \textbf{Leaking Instructions} \\
            \midrule
            Intel Core i5-2520M           &   0x2f       &   Sandy Bridge   &  2011     &     Yes     & \instr{rdtsc}, \instr{rdtscp}  \\ %
            Intel Core i5-3230M           &   0x21       &   Ivy Bridge     &  2013     &     Yes     & \instr{rdtsc}, \instr{rdtscp}, \instr{sldt}  \\ %
            Intel Core i3-4160T           &   0x28       &   Haswell        &  2014     &     Yes     & \instr{rdfsbase}, \instr{rdgsbase}   \\ %
            Intel Core i3-5010U           &   0x2f       &   Broadwell      &  2015     &     Yes     & \makecell[l]{\instr{rdfsbase}, \instr{rdgsbase},\\ \instr{rdtsc} (U), \instr{rdtscp} (U)}   \\ %
            Intel Atom x5-Z8350           &   0x411      &   Cherry Trail   &  2016     &     Yes     & \instr{rdpmc} \\ %
            Intel Celeron N3550           &   0x28       &   Apollo Lake    &  2016     &     No     & \instr{rdpmc} \\ %
            Intel Celeron J4005           &   0x3c       &   Gemini Lake    &  2017     &     Yes     & \instr{rdpmc} \\ %
            Intel Core i3-7100T           &   0xf0       &   Kaby Lake      &  2017     &     Yes     & \instr{rdfsbase}, \instr{rdgsbase}   \\ %
            Intel Core i3-1005G1          &   0xb2       &   Ice Lake       &  2019     &     No     & --                                   \\ %
            Intel Core i7-10510U          &   0xf0       &   Comet Lake     &  2019     &     No     & \instr{rdfsbase}, \instr{rdgsbase}   \\ %
            Intel Core i7-1185G7          &   0xa4       &   Tiger Lake     &  2020     &     No     & --   \\ %
            Intel Celeron N4500           & 0x240000023~  &   Jasper Lake    &  2021     &     No     & \instr{rdfsbase} (ZF), \instr{rdgsbase} (ZF), \instr{sldt} (ZF) \\ %
            Intel Core i9-12900K          & 0x22         &   Alder Lake     &  2021     &     No     & --    \\ %
            Intel Atom x6425E             & 0x17         &   Elkhart Lake   &  2021     &     No     & --    \\ %
            AMD GX-415GA                  & 0x700010f    &   Jaguar         &  2013     &     No     & --   \\ %
            AMD Ryzen 5 2500U             & 0x810100b    &   Zen            &  2017     &     No     & \instr{rdfsbase}, \instr{rdgsbase}   \\ %
            AMD Ryzen 5 3550H             & 0x8108102    &   Zen+           &  2019     &     No     & \instr{rdfsbase}, \instr{rdgsbase}   \\ %
            AMD Epyc 7252                 & 0x8301055    &   Rome           &  2019     &     No     & \instr{rdfsbase}, \instr{rdgsbase}, \instr{str} (ZF) \\ %
            AMD Ryzen 9 6900HX~            & 0xa404102    &   Zen 3+         &  2022     &     No     & \instr{rdfsbase}, \instr{rdgsbase}   \\ %
            \bottomrule
        \end{tabular}
        }
\end{table}

\section{Attack Primitive}\label{sec:primitive}
In this section, we introduce our attack primitive \PerfAttack.
\PerfAttack exploits \meltdownthreea to leak performance-counter values using \instr{rdpmc} to infer side-channel information about program executions.

\subsection{Threat Model}\label{subsec:threat-model}

We assume an unprivileged attacker with native code execution. %
We further assume bug-free victim software, \eg the absence of memory corruption or logical vulnerabilities.
However, our attacker model relies on side-channel vulnerabilities, \ie we assume secret-dependent control or data flow in the victim application.
Even though our attacks are, in theory, mountable from inside virtual machines, we did not explicitly test this, and attackers could only target victims inside their own virtual machine and not the hypervisor or other virtual machines.
While this weakens the attack surface, intra-VM attacks are still a realistic scenario, \eg in container-based environments.
We target only Intel and AMD CPUs in this work.
Note that \meltdownthreea is also exploitable on Arm \cite{ARMSpecAnalysis_whitepaper} but we consider further architectures out of scope for the experiments conducted in this paper and only discuss them in \Cref{sec:discussion}.

\subsection{\PerfAttack}\label{sec:perfattack}

\begin{figure}[t]
    \centering
    \input{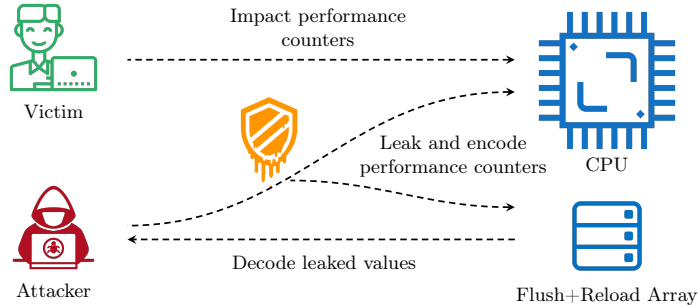}
    \caption{\meltdownthreea leaking system registers, such as performance counters.}
    \label{fig:meltdown3a-overview}
\end{figure}

The \PerfAttack attack primitive relies on \meltdownthreea.
We use \meltdownthreea to infer side-channel information about a victim program. 
Based on our systematic analysis using \AnalysisTool, and the publicly-available information regarding \meltdownthreea by Intel~\cite{IntelrogueSystemRegRead2018AffectedRegisters}, we build our attack primitive on top of \instr{rdpmc}.
\instr{rdpmc} provides a generic but privileged interface to performance counters.
Access to these performance counters leaks information about the program execution that can be exploited for side-channel attacks~\cite{Uhsadel2008,Bhattacharya2015,Dixon2017}. 

\parhead{Attack Overview}
\PerfAttack relies on \meltdownthreea to leak the content of a performance counter. 
We assume that the system already has a performance counter programmed. 
This is the case if the system uses performance counters for attack detection, as suggested by previous work~\cite{Herath2015,Payer2016,Irazoqui2016mascat,Zhang2016CloudRadar}. 
For example, Cloudflare relies on performance counters to detect Spectre attacks~\cite{Varda2021DPI}. 
An attacker leaks the performance-counter values by encoding the transiently-read return value of the \instr{rdpmc} instruction into the microarchitecture and recovers it using a side channel. 

\parhead{Implementation}
In line with previous Meltdown-type attacks~\cite{Lipp2018meltdown,Kocher2019,Schwarz2019ZL,Vanbulck2018foreshadow,VanSchaik2019RIDL,Moghimi2020medusa,Canella2019A}, we use the CPU cache to encode the transiently-leaked values and \FlushReload as the covert channel to make the values architecturally visible.
We support leakage of 1 to 4 bytes per \instr{rdpmc} invocation by encoding each byte into the cache state of an array consisting of \num{256} pages.
The more data is encoded into the microarchitecture, the better the resolution of the underlying performance counter value.
However, this also leads to a slower decoding phase, as more \FlushReload attacks are required. 
For leaking a single byte, at most \num{256} \FlushReload attacks are necessary, while for leaking 4 bytes, at most \num{1024} \FlushReload attacks are necessary.
We evaluate this trade-off in \Cref{sec:eval}.

\section{Evaluation} \label{sec:eval}

In this section, we evaluate the attack primitive \PerfAttack which is based on \meltdownthreea.
All evaluations use our proof-of-concept implementation on an \LabOneCPU running \LabOneOS with \LabOneKernel.

\begin{figure}[t]
    \centering
    \resizebox{\hsize}{!}{
    \begin{tikzpicture}
\begin{axis}[
style={font=\footnotesize},
xlabel={Experiment Iteration},
ymode=log,
ylabel={\parbox{2cm}{\centering Time Required \smallbreak [in CPU cycles]}},
width=\hsize,
xmin=0,
ymax=70000000,
xmax=99,
height=3.5cm,
legend style={at={(1,0.933)},anchor=east},
legend columns=4
]

\addplot+[blue, very thick,no marks] table[x=leak-no,y=1byte,col sep=comma] {data/leakage_time.csv};
\addplot+[green, very thick,no marks] table[x=leak-no,y=2byte,col sep=comma] {data/leakage_time.csv};
\addplot+[orange, very thick,no marks] table[x=leak-no,y=3byte,col sep=comma] {data/leakage_time.csv};
\addplot+[red, very thick,no marks] table[x=leak-no,y=4byte,col sep=comma] {data/leakage_time.csv};

\legend{1 Byte, 2 Bytes, 3 Bytes, 4 Bytes};
\end{axis}
\end{tikzpicture}
    }
    \caption{\PerfAttack: CPU cycles needed to leak, i.e., access, encode, and decode, $n$ bytes of a performance counter by attacking \instr{rdpmc}. The y-axis shows the CPU cycles required for each repetition of the experiment (x-axis).}
    \label{fig:perfattack-leaking-time} 
\end{figure}
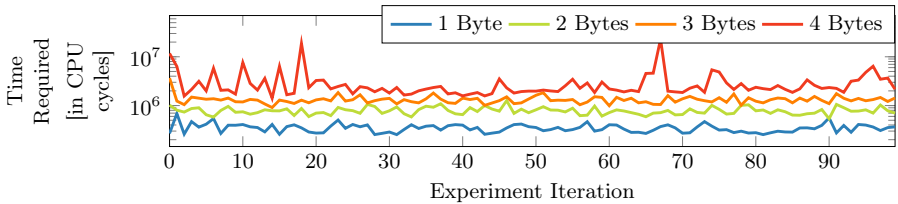

\begin{figure}[t]
\resizebox{\hsize}{!}{
        \begin{tikzpicture}
\begin{axis}[
style={font=\footnotesize},
xlabel={Experiment Iteration},
ylabel={\parbox{2cm}{\centering Time Required [in CPU cycles]}},
width=\hsize,
xmin=0,
xmax=100,
height=3cm,
]

\addplot+[very thick,no marks] table[x=leak-no,y=value,col sep=comma] {data/effective_leak_time.csv};
\end{axis}
\end{tikzpicture}
}
        \caption{\PerfAttack: CPU cycles needed to transiently encode 4 bytes of the CPU timestamp counter. The y-axis shows the CPU cycles required for each repetition of the experiment (x-axis).
        }
        \label{fig:perfattack-rdtsc-time}
\end{figure}
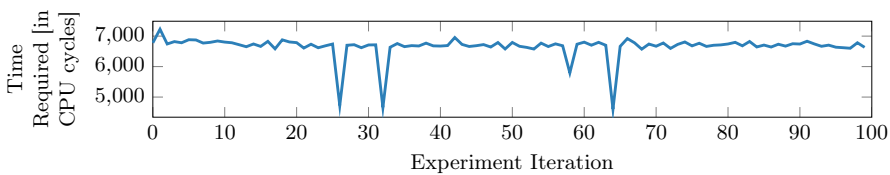

The most important property in our evaluation is the temporal resolution of \PerfAttack, \ie the time between two measurements.
This property reflects how fine-grained the information can be leaked by the exploit.
We evaluate the time it takes to leak $n$ bytes of a system register.
This measurement directly gives us the temporal resolution of the attack.
We observe that the implementation leaks 1 byte of a system register in, on average, \SIx{348257} cycles ($n$ = \SIx{100}).
\Cref{fig:perfattack-leaking-time} summarizes the time an attacker needs to leak the content of a performance counter when leaking $n$ bytes within one transient window.
We emphasize that this is a good indication of the theoretical performance of this attack, as an attacker can likely mount exploits by only leaking parts of the system register.
We also require only partial leakage for our attacks discussed in \Cref{sec:case}.
Note that the temporal resolution mostly affects the execution time of an attack but does not prevent an attack.
An attacker can often compensate for a lower temporal resolution by averaging over repeated measurements~\cite{Lipp2020Platypus}.

Still, whereas our complete attack primitive takes millions of CPU cycles for one iteration (\cf \Cref{fig:perfattack-leaking-time}), the actual time spent encoding multiple bytes of a system register is significantly shorter.
While the time needed to leak $n$ bytes of a performance counter, \ie the attack's temporal resolution, is important for repeated measurements, another critical metric is the time that an attacker needs to encode a value in the CPU cache.
This metric is especially important for event-driven attack scenario, \ie whenever the attacker wants to take a measurement after a certain event has happened.
To evaluate the time it takes to encode a value, we record the time needed to encode the value of the timestamp register over 100 runs.
\Cref{fig:perfattack-rdtsc-time} shows the results.
We observe that the average time between the faulting access and the first subsequent attacker-controlled instruction when encoding 4 bytes simultaneously is \num{6655} cycles.
Whereas the effective blindspot of our attack is higher, this time yields the offset between an event triggering a measurement in the attacker code and the measurement itself.

\section{Case Studies} \label{sec:case}
In this section, we introduce 4 case studies demonstrating \PerfAttack.
We demonstrate a Spectre proof-of-concept (PoC) (\Cref{subsec:spectre-counterleak}) and break KASLR by monitoring the behavior of page walks (\Cref{subsec:kaslr-break}).
To demonstrate that our side channels re-enable mitigated attacks, we leak a 2048-bit RSA private key from a square-and-multiply implementation found in MbedTLS using \PerfAttack (\Cref{subsec:attacking-cryptographic-keys-via-pmc-leakage}).
Lastly, we show that we can break the branch-shadowing mitigation proposed by Lee~\etal\cite{Lee2017Inferring} using \PerfAttack (\Cref{subsec:breaking-branch-shadowing-mitigation}).
\subsection{Spectre with \PerfAttack}\label{subsec:spectre-counterleak}

In this case study, we demonstrate a Spectre-type attack~\cite{Kocher2019,Canella2019A} with our \PerfAttack primitive to leak otherwise inaccessible data.
We build a Spectre-PHT~\cite{Kocher2019,Canella2019A} PoC with a performance counter as covert channel. 

\parhead{Target Performance Counter}
We target a performance counter that tracks speculative events~\cite{Qiu2022PMUSpill}, such as \code{CYCLES\_DIV\_BUSY.ALL} and assume that it is either activated or can be enabled by the attacker.
Note that depending on the victim's code, the discussed attack can also be mounted with a different performance counter.
The only requirement is that the accessed secret can be encoded in branches that can be distinguished based on any performance counter.

\parhead{Attack Overview}
We attack a Spectre gadget of the form 
\begin{lstlisting}
if (i >= 0 && i < array_size) {
    int tmp = (array[i] >> offset);
    if ((tmp & 1))  x / y; 
}\end{lstlisting}
The attacker controls the variables \code{i} and \code{offset}.
Note that even though the inner \texttt{if} branch is only doing an operation that should not result in any state change, it still affects related performance counters and hence suffices to enable our attack.
The attacker starts by mistraining the outer \texttt{if} branch such that its subsequent execution is misspeculated to be taken.
The simplest way to achieve this is by in-place mistraining~\cite{Canella2019A}, \ie executing the branch multiple times with \code{i} being a valid offset for the array. 
As a baseline, the attacker leaks the value of the performance counter \code{CYCLES\_DIV\_BUSY.ALL} using \PerfAttack.
This performance counter keeps track of the number of cycles the CPU's divider units are used.
The attacker executes the victim function with an index \code{i} that is outside the bounds of the array and corresponds to the targeted memory address. %
Afterward, the attacker again leaks the performance counter of \code{CYCLES\_DIV\_BUSY.ALL} using \PerfAttack and subtracts the previously leaked value.
As the divider is only used when the inner \texttt{if} branch is (speculatively) taken, the delta is slightly higher if the transiently-accessed bit is `1'.

\parhead{Results}
We measure each bit 50 times and set a threshold on the median to distinguish between `1' and `0' bits based on the value of the performance counter. %
Our PoC achieves a leakage rate of \SI{66.7}{\bit/\second} with an accuracy of \SI{99.6}{\percent}.
While not the fastest covert channel, we argue that it is still fast enough to pose a threat when such an attack is mounted.

\parhead{Comparison to Similar Attacks}
Our attack only relies on a control flow that is distinguishable by observing performance counters.
Common covert channels used in Spectre-type attacks require cache accesses to encode data from transient execution~\cite{Kocher2019,Koruyeh2018spectre5,Hetterich2022Branch,Maisuradze2018spectre5}.
Finding such code paths that can be exploited by Spectre-type attacks, also referred to as Spectre gadgets, is a challenging task.
While our attack is limited to a CPU vulnerable to \PerfAttack and providing a usable performance counter, it can use both traditional Spectre gadgets and novel types of gadgets.
Hence, with the combination of Spectre and \PerfAttack, the number of potential gadgets increases. 

\subsection{Breaking KASLR with \PerfAttack}\label{subsec:kaslr-break}
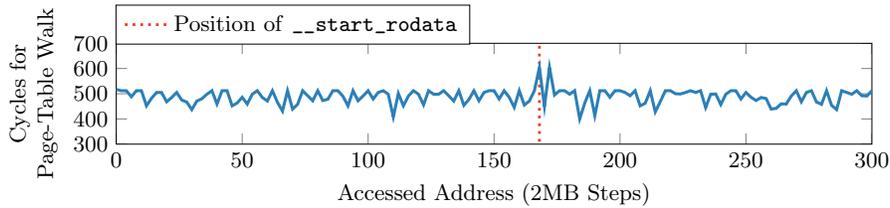
\begin{figure}[t]
    \centering
    \resizebox{\hsize}{!}{
    \begin{tikzpicture}
\begin{axis}[
style={font=\footnotesize},
xlabel={Accessed Address (2MB Steps)},
ylabel={\parbox{2.5cm}{\centering Cycles for Page-Table Walk}},
width=\hsize,
ymax=700,
ymin=300,
xmax=300,
xmin=0,
height=3cm,
legend style={at={(-0.001,1.185)},anchor=west}
]

\addplot+[mark=none,line width=1.1pt,dotted,color=red] coordinates {(168, 310) (168, 800)};
\addlegendentry{Position of \code{\_\_start\_rodata}}
\addplot+[very thick,no marks, color=blue, forget plot] table[x=offset,y=cycles,col sep=comma] {data/kaslr.csv};
\end{axis}
\end{tikzpicture}
    }
    \caption{The leaked values of \code{DTLB\_LOAD\_MISSES.WALK\_COMPLETED\_2M\_4M} when iterating over the potential locations where the Linux kernel could be mapped. The page-table walk needs longer when the address is the actual start of the kernel, i.e., the position of the kernel symbol \code{\_\_start\_rodata}.}
    \label{fig:kaslr-break}
\end{figure}

We demonstrate that unprivileged access to performance counters breaks Kernel Space Address Layout Randomization (KASLR).
KASLR randomizes the base address of the operating system kernel upon booting.
As precise knowledge of the memory layout is a requirement for many attacks, KASLR adds an additional barrier that attackers have to overcome for a successful kernel exploit.
We show that we can derandomize the location of the Linux kernel on an \LabFifteenCPU running \LabFifteenOS with \LabFifteenKernel and thus bypass KASLR.

\parhead{Target Performance Counter}
We target a performance counter influenced by page-table walks, such as \code{DTLB\_LOAD\_MISSES.WALK\_COMPLETED\_2M\_4M} and assume that is already programmed or can be programmed by the attacker.
A scenario in which this is the case is if the system is protected using the approach of Wang~\etal\cite{Wang2020hybrid}.

\parhead{Attack Overview}
For derandomizing the kernel location, we rely on the property that non-present pages are not stored in the TLB~\cite{Canella2020kaslr}.
Thus, a memory load request to a non-present page always leads to a page-table walk, whereas a memory load request to a present page leads to a TLB hit, resulting in no page-table walk if the page was recently accessed.
The attack iterates over each potential location of the kernel and accesses it.
The resulting fault caused by the access is suppressed using speculative execution, TSX transactions, or fault handling.
For each memory access, the attacker leaks the performance counter \code{DTLB\_LOAD\_MISSES.WALK\_COMPLETED\_2M\_4M}, or an alternative one correlating to the number of or the cycles spent for page-table walks, using \PerfAttack.
Based on the leaked value, the attacker can observe whether a memory page is present and was recently accessed.
The first page of the kernel's \code{.rodata} section is frequently accessed.
Thus, the first address showing an abnormal timing difference is the location of the kernel symbol \code{\_\_start\_rodata}.
Note that a more advanced version of this attack can also be used to actively monitor the access to kernel memory pages, similar to the work of Schwarz~\etal\cite{Schwarz2019STL}.

\parhead{Results}
\Cref{fig:kaslr-break} shows the cycle difference iterating over the kernel address space.
The kernel location is easily distinguishable from non-present pages due to the change in cycles spent for page table walks.
We tested our KASLR break on an \LabOneCPU running \LabOneOS with \LabOneKernel observing a success rate of \SI{98}{\percent} (n=100) and a median execution time of \SI{4.7}{\second}.

\subsection{Attacking RSA with \PerfAttack}\label{subsec:attacking-cryptographic-keys-via-pmc-leakage}

In this case study, we attack the RSA implementation based on the MbedTLS version 1.3.10 running on an \LabOneCPU with \LabOneOS and \LabOneKernel.
This MbedTLS version implements RSA by using a window-based square-and-multiply algorithm.
We configure the window size to 1. %
Previous work~\cite{Liu2015Last} showed that all window sizes are vulnerable if window size 1 is vulnerable.
While such square-and-multiply implementations are known to be vulnerable to side-channel attacks, we choose this target as it is a common target for related attacks~\cite{Schwarz2017MGX,huo2020bluethunder,Garcia2020Certified,Lee2017Inferring,Xiao2017stacco}.
Hence, we ease comparison with other side-channel attacks. 

\parhead{Target Performance Counter}
We target the performance counter \code{BR\_INST\_RETIRED.NEAR\_TAKEN} and assume that it is either already programmed or can be programmed by the attacker.
This performance counter keeps track of the number of taken near-branch instructions.
An example for a realistic scenario in which this performance counter would be programmed is a system protected by the rootkit detection of Singh~\etal\cite{Singh2017rootkit}.

\parhead{Attack Overview}
The victim application consists of a branch only taken when the currently-processed secret bit is `1'.
Thus, the secret bit correlates with the number of branches taken.
The attacker gains oracle access to the signing routine of the application to sign arbitrary messages. %
We assume that the attacker and victim are synchronized, \ie the attacker either knows when the victim processes each iteration of the exponentiation loop, or the attacker can influence this by, \eg interrupting the victim.
During the execution of the victim, the attacker repeatedly leaks the value of the performance counter and, thereby, the number of branches taken.
The attacker leaks the performance counter once per key bit.
Afterward, the attacker stores the delta of two consecutive performance counter leaks, \ie the approximation of the victim's taken branches for the processing of a specific secret bit.
The attacker repeats this procedure for the decryption of \SIx{10000} different messages, averaging out the noise of branches taken by \PerfAttack itself and the unrelated branches of the victim application.

\begin{figure}[t]
    \centering
    \resizebox{\hsize}{!}{
    \begin{tikzpicture}
\begin{axis}[
style={font=\footnotesize},
xlabel={Secret Bits},
ylabel={\parbox{2cm}{\centering Observed Number of Branches}},
x tick style={draw=none},
xticklabels={,,},
width=\hsize,
xmin=1000,
ymin=5300,
xmax=1019.5,
height=3cm,
]
\draw[red] (axis cs: 1000,5500) to (axis cs: 1020,5500);
\node at (axis cs: 1000.1,5330) {1};
\node at (axis cs: 1001.0,5330) {1};
\node at (axis cs: 1002.0,5330) {1};
\node at (axis cs: 1003.0,5330) {0};
\node at (axis cs: 1004.0,5330) {1};
\node at (axis cs: 1005.0,5330) {1};
\node at (axis cs: 1006.0,5330) {1};
\node at (axis cs: 1007.0,5330) {0};
\node at (axis cs: 1008.0,5330) {1};
\node at (axis cs: 1009.0,5330) {1};
\node at (axis cs: 1010.0,5330) {1};
\node at (axis cs: 1011.0,5330) {0};
\node at (axis cs: 1012.0,5330) {1};
\node at (axis cs: 1013.0,5330) {1};
\node at (axis cs: 1014.0,5330) {0};
\node at (axis cs: 1015.0,5330) {0};
\node at (axis cs: 1016.0,5330) {1};
\node at (axis cs: 1017.0,5330) {0};
\node at (axis cs: 1018.0,5330) {0};
\node at (axis cs: 1019.0,5330) {0};

\addplot+[very thick,no marks] table[x=idx,y=pmc-median,col sep=comma] {data/square-multiply.csv};
\end{axis}
\end{tikzpicture}
    }
    \caption{The leaked value of the performance counter \code{BR\_INST\_RETIRED.NEAR\_TAKEN} and its correlation to the secret bits of the exponent.}
    \label{fig:perf-rsa-leak}
\end{figure}
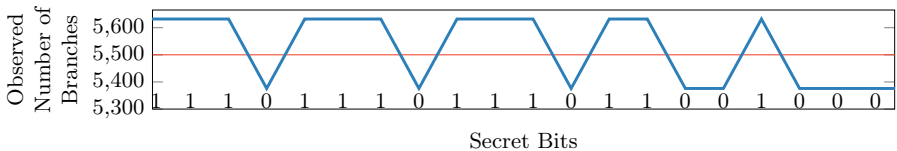

\parhead{Results}
By averaging over \SIx{10000} traces, we extract a clear indication of the secret bits.
\Cref{fig:perf-rsa-leak} visualizes the correlation between the number of branches taken and the secret bits.
Using a simple threshold, we recover \SI{99.9}{\percent} of the 2048-bit RSA keys ($n$ = 10) in around \SI{15}{\minute}.
Compared to previous work, there are both faster attacks requiring fewer encryptions~\cite{Aciicmez2007icache,Liu2015Last} and attacks requiring a similar number of decryptions or more time to execute~\cite{Yarom2017cachebleed,Zhang2012}.
We conclude that \PerfAttack yields a strong primitive for leaking secrets from, for example, cryptographic implementations.

\subsection{Breaking Zigzagger with \PerfAttack}\label{subsec:breaking-branch-shadowing-mitigation}

In this case study, we explore how \PerfAttack breaks the Zigzagger branch-shadowing mitigation.
Branch-shadowing attacks exploit the shared branch history between processes, allowing attackers to reason about the direction of a branch.
For example, Lee~\etal\cite{Lee2017Inferring} demonstrate that a branch-shadowing attack can leak confidential data from Intel SGX enclaves.
To prevent branch-shadowing attacks, Lee~\etal\cite{Lee2017Inferring} proposed a software mitigation called Zigzagger.
Zigzagger replaces a set of branches with a single indirect branch.
Thus, the attacker can only infer whether the branch was executed but cannot infer the branch direction anymore.
To compute the address of the indirect jump, additional conditional-move instructions are used.
In line with Gerlach~\etal\cite{Gerlach2023SecurityRISC}, we exploit the number of retired instructions to break the Zigzagger mitigation.
While Gerlach~\etal used an architectural interface to this information, we show that we can recover the same information using \PerfAttack.
This information allows an attacker again to distinguish the branches taken by the victim.

\parhead{Target Performance Counter}
We target the \code{INSTR\_RETIRED} performance counter that is either already programmed or can be programmed by the attacker.
A realistic scenario for this would be if the defense approach of Wang~\etal\cite{Wang2020hybrid} is in use on the system.

\parhead{Attack Overview}
The victim process contains secret-dependent branches and is hardened against branch-shadowing attacks using Zigzagger~\cite{Lee2017Inferring}.
The attacker leaks the \code{INSTR\_RETIRED} performance counter before and after the Zigzagger-hardened victim executes.
The delta between these measurements yields the number of retired instructions.
The attacker correlates this number with a baseline measurement for all branches.

\parhead{Results}
For the case study, we use an \LabOneCPU running \LabOneOS with \LabOneKernel.
For each of the 3 different possible arguments of the sample function, there is a unique number of retired instructions after the Zigzagger modification was applied.
Hence, by observing the number of retired instructions, an attacker can directly infer the arguments. 
We observe a success rate of \SI{100}{\percent} using \SIx{10000} recorded measurements.

\section{Countermeasures} \label{sec:countermeasures}
In this section, we discuss countermeasures against \PerfAttack and \meltdownthreea.
The fundamental problem is that an unprivileged attacker can transiently access the metadata of an application in the form of performance counters.
The exploited vulnerability is rooted deep inside the CPU.
As the information stem from a CPU register, no software is involved.
Nevertheless, operating systems can still defend against the impact of the attack whereas the victim application itself can be hardened against the attack.

\parhead{Firmware}
Several CPUs received microcode updates to prevent the leakage of system registers~\cite{rogueSystemRegRead2018}.
While CPU vendors do not disclose internals of these updates, it is likely that a similar patch can also mitigate the remaining leakage. %
Thus, the most efficient and effective mitigation is likely via microcode updates.

\parhead{Kernel}
\PerfAttack fundamentally relies on performance counters that are either already programmed or that can be programmed by an attacker-accessible API.
A common scenario for this are performance-counter-based detection approaches~\cite{Zhang2013,Irazoqui2018mascat,Payer2016,Zhang2016CloudRadar,Chiappetta2015,Mushtaq2020,Mushtaq2018nightswatch,Wang2020hybrid,Wang2020comprehensive,Zhang2021see}.
As the absence of programmed or programmable performance counter prevents \PerfAttack, a carefully designed system that does not use performance counters at all or only in the absense of untrusted parties and code can also prevent the exploitation of \PerfAttack.
As performance counters and their programming requires kernel privileges, the kernel could, in theory, completely prevent the programming of performance counters.
However, this decision comes with the drawback that it would break existing software like the performance-counter-based detection approaches or monitoring utilities such as \code{perf}.
In contrast, an operating system can prevent attacks on KASLR without breaking existing software. 
Canella~\etal\cite{Canella2020kaslr} proposed mapping dummy pages in the kernel such that all kernel addresses are mapped. 
Consequently, an attacker cannot infer the real location of the kernel. 

\parhead{Userspace Software}
As \PerfAttack is a side-channel attack, it is fundamentally limited to leaking data from an application with secret-dependent branches or data-flow edges.
However, an application can generally be implemented without any secret-dependent accesses~\cite{Intel2020Guidelines}.
Applications implemented in such a way are not susceptible to \PerfAttack.
Especially for cryptographic algorithms, such implementations are state-of-the-art.

\section{Discussion} \label{sec:discussion}
In this section, we discuss related work.
Furthermore, we show how the presented attack primitive behaves on different operating systems and architectures.
As the building blocks of \PerfAttack are OS-agnostic and also exist on other architectures, we assume that similar attacks are also possible there.

\subsection{Related Work}
In 2018, Intel and Arm disclosed the vulnerability and assigned it CVE-2018-3640~\cite{IntelRSRRadvisory,ARMSpecAnalysis_whitepaper}.
While Intel released a security advisory and added a new category to their list of CPUs affected by vulnerabilities~\cite{IntelRSRRadvisory,Intel2023Affected}, Arm added a section about the vulnerability in their Cache-Speculation Side-Channel whitepaper~\cite{ARMSpecAnalysis_whitepaper}.
Our work builds on this initial disclosure by analyzing the leakage of different system registers on 19 CPUs with applied vendor mitigations.
We further demonstrate that it is still possible to exploit \meltdownthreea in different scenarios.

While we focus our work on \meltdownthreea, Canella~\etal\cite{Canella2019A} analyzed the landscape of transient-execution attacks with a broader focus.
Furthermore, they first introduced the split into Meltdown- and Spectre-type attacks.
In contrast, our work focuses on the specific variant \meltdownthreea and analyzes further details about it, including how widespread the issue itself is.

Attacks exploiting performance counters have been shown when the interface was accessible to unprivileged users. 
In 2008, Uhsadel~\etal\cite{Uhsadel2008} first exploited performance counters to leak information about the CPU caches.
With information similar to a cache attack, they showed that the information can be exploited to recover confidential values from a victim program.
They also demonstrated their attack on an OpenSSL AES implementation.
Bhattacharya~\etal\cite{Bhattacharya2015} further demonstrated that performance counters expose even more information than just the cache state and thus allow reasoning about the branch-predictor state.
Their work discusses an exploit on a square-and-multiply implementation of RSA using the Montgomery-ladder algorithm.
Since then, the access to performance counters is privileged by default, preventing these attacks on modern systems~\cite{Dixon2017}.
Dixon~\etal\cite{Dixon2017} further stresses the importance of disabling unprivileged access to performance counters by showing that it allows derandomizing the kernel location.
Gerlach~\etal\cite{Gerlach2023SecurityRISC} exploit the unprivileged access to performance counters on RISC-V CPUs to break KASLR, leak the presence of inaccessible files, and detect interrupts. 
Our work mainly differs from these previous ones by demonstrating these and similar attacks on modern systems where performance-counter access is restricted to privileged users.

\subsection{Other OS and Architectures}
The underlying effects exploited in this paper are OS-agnostic.
While this paper targets Linux, we do not require any Linux-specific functionality.
\PerfAttack interacts with the hardware directly without requiring any OS support.
If any application legitimately enables performance counters, they can be leaked.

\PerfAttack requires systems that are vulnerable to \meltdownthreea. 
While \meltdownthreea was also shown on Arm CPUs~\cite{ARMSpecAnalysis_whitepaper}, we leave it for future work to systematically analyze Arm CPUs for their \meltdownthreea attack surface. %
Nevertheless, as all strict requirements for \PerfAttack are also given on Arm CPUs, we suspect that the issue also affects these systems.

\section{Conclusion} \label{sec:conclusion}
In this paper, we analyzed the attack surface of \meltdownthreea.
For this, we developed an automated approach using \AnalysisTool (open-sourced on GitHub) to analyze 19 Intel and AMD CPUs based on different microarchitectures.
In our analysis, we observe that the privileged system registers that can be leaked by \meltdownthreea differ from CPU to CPU.
Furthermore, we observe that the FS and GS segment base registers can be leaked even on recent AMD CPUs (Zen 3+).
We further show that our attack primitive \PerfAttack can exploit side-channel information by leaking the values of performance counters using \meltdownthreea.
We demonstrated \PerfAttack in 4 different case studies.
We showed that the primitive allows us to break KASLR by monitoring the page-table walker and can break the Zigzagger branch-shadowing mitigation~\cite{Lee2017Inferring}.
Additionally, we demonstrated the applicability of \PerfAttack as a flexible covert channel for Spectre attacks and leaked a 2048 bit RSA key from a square-and-multiply implementation in MbedTLS, verifying that our primitive reenables previously mitigated attacks. 
In conclusion, our work shows that \meltdownthreea should not be underestimated and still poses a threat to modern and fully patched systems.

\section*{Acknowledgment}
We want to thank our anonymous reviewers for their comments and suggestions.
We also want to thank Leon Trampert and Niklas Flentje for providing their help with running the experiments. 
This work was partly supported by the Semiconductor Research Corporation (SRC) Hardware Security Program (HWS).

\bibliographystyle{IEEEtranS}
\bibliography{main}
\FloatBarrier

\end{document}